\documentclass[aip,reprint]{revtex4-1}
\usepackage{graphicx}
\usepackage{soul,color}

\begin{document}

\draft 

\title{Magnetoelectric coupling in a frustrated spinel studied using high-field scanning probe microscopy} 
\author{L. Rossi}
\affiliation{High Field Magnet Laboratory (HFML-EMFL) and Institute for Molecules and Materials, Radboud University, Toernooiveld 7, 6525 ED Nijmegen, The Netherlands}

\author{D. Br\"uning}
\affiliation{Physikalisches Institut, Universit\"at zu K\"oln, Z\"ulpicher Str. 77, 50937 K\"oln, Germany}

\author{H. Ueda}
\affiliation{Department of Chemistry, Graduate School of Science, Kyoto University, Kyoto 606-8502, Japan}

\author{Y. Skourski}
\affiliation{Hochfeld-Magnetlabor Dresden (HLD-EMFL), Helmholtz-Zentrum Dresden-Rossendorf, D-01314 Dresden, Germany}

\author{T. Lorenz}
\affiliation{Physikalisches Institut, Universit\"at zu K\"oln, Z\"ulpicher Str. 77, 50937 K\"oln, Germany}

\author{B. Bryant}
\email{ben.e.bryant@gmail.com}
\thanks{Current address: Oxford Instruments NanoScience, OX13 5QX, United Kingdom}
\affiliation{High Field Magnet Laboratory (HFML-EMFL) and Institute for Molecules and Materials, Radboud University, Toernooiveld 7, 6525 ED Nijmegen, The Netherlands}

\date{\today}

\begin{abstract}
Below its Ne\'el temperature, the frustrated magnet CdCr$_2$O$_4$ exhibits an antiferromagnetic  spin-spiral ground state. Such states can give rise to a sizable magnetoelectric coupling. In this report, we measure the electric polarization induced in single-crystalline CdCr$_2$O$_4$ by large applied magnetic field. Because the detection of a macroscopic polarization is hindered by the structural domains in the tetragonal spin-spiral phase, we have pioneered an alternative method of measuring polarization induced by high magnetic fields, using electrostatic force microscopy. This method enables us to measure polarization from nanometer sized areas of the sample surface, as well as imaging how charge inhomogeneities change with magnetic field.
\end{abstract}

\pacs{}

\maketitle

Materials with strong magneto--electric coupling hold the promise of revolutionizing information storage through electric field control of magnetism. However, such coupling is often weak, since ferroelectricity and magnetism derive from breaking of different symmetries \cite{Cheong2007}. Non--collinear magnetic ground states, such as spin--spirals, often found in frustrated magnets, can provide a route to strong magneto--electric coupling \cite{Tokura2010,Arima2011}. 

In the chromium spinels ACr$_{2}$X$_4$ (A = Zn, Cd, Hg: X = O, S, Se), the Cr--ions form a pyrochlore lattice of corner--sharing tetrahedra, leading to spin frustration in the antiferromagnetic members of the family. Many of these compounds present a spin--spiral ground state, including ZnCr$_2$Se$_4$ \cite{Murakawa2008} and the oxides ACr$_{2}$O$_4$ \cite{Matsuda2010}. The oxide compounds also exhibit a half--magnetization ``plateau'' state under applied magnetic field \cite{Ueda2005,Ueda2006,Miyata2012}. In the antiferromagnetic compound CdCr$_2$O$_4$, there is a transition to an incommensurate spin--spiral below the Ne\'el temperature $T_N$ of 7.8~K \cite{Matsuda2007}; at low temperature, the spin--spiral disappears at a magnetic field of 28~T, where the material enters the plateau state \cite{Ueda2005}. The plateau state has a spin structure in which three of the spins in each tetrahedron point ``up'', and one points ``down'' \cite{Penc2004, Ueda2005, Ueda2006, Shannon2006, Shannon2010}.

In this paper, we present the generation of electric polarization by applied magnetic field in the spin--spiral ground state of the frustrated spinel CdCr$_2$O$_4$. We measure the field--induced electric polarization both via a standard polarization current (pyrocurrent) technique, and by measuring the contact potential difference (CPD) between the sample surface and the tip of a scanning probe microscope, in a static magnetic field up to 30~T generated by a Bitter magnet. In order to perform these measurements, we employed a high--field scanning probe microscope (HF--SPM) \cite{Rossi2018}. The SPM technique has the advantage of being able to measure the surface polarization within an area much smaller than the size of the tetragonal structural domains, circumventing the problems these domains present for traditional polarization measurements.

\begin{figure}[b]
	\begin{centering}
    \includegraphics[scale=1]{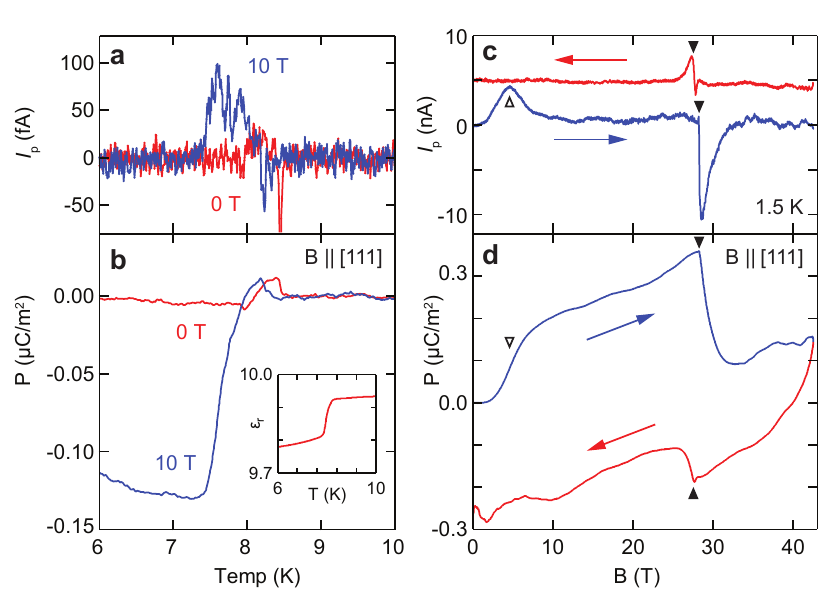}
	\caption{Temperature dependence of (a) polarization current \emph{I}$_P$ and (b) electric polarization $P$, of CdCr$_2$O$_4$ at zero field and 10~T, measured on warming. Inset: corresponding change in relative permittivity $\epsilon_r$. Magnetic field dependence of (c) \emph{I}$_P$ and (d) $P$, recorded at up to 42.5~T in a pulsed magnet at 1.5~K. The spin flop is indicated by open triangles, the transition to the plateau state by filled triangles. P is measured parallel to the B field: no poling voltage was applied.}
	\label{Polarization}
	\end{centering}
\end{figure}

\begin{figure}[b]
	\begin{centering}
    \includegraphics[scale=1]{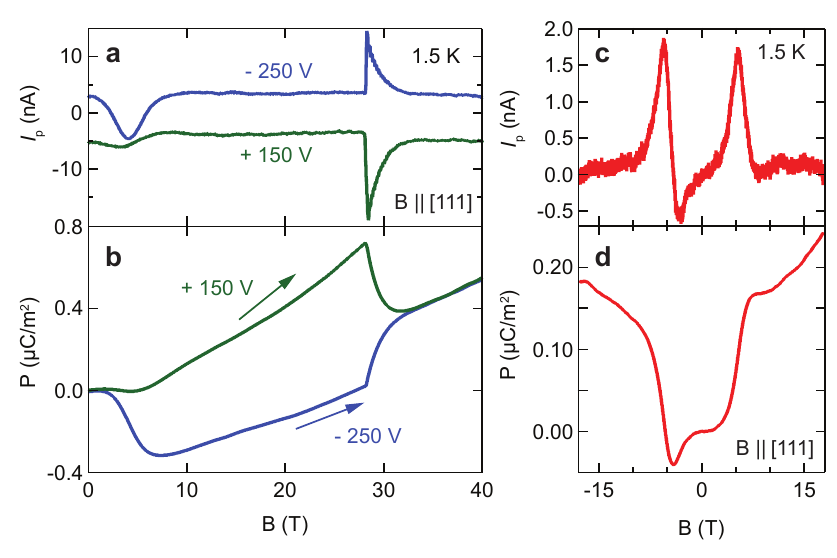}
	\caption{Dependence of polarization on poling voltage and field polarity. Magnetic field dependence of (a) \emph{I}$_P$ and (b) $P$, recorded at up to 40~T in a pulsed magnet at 1.5~K, after cooling under the poling voltage indicated. (c) and (d) show the effect on \emph{I}$_P$ and $P$ of changing the magnetic field polarity, no inversion of electric polarization is observed. P is measured parallel to the B field.}
	\label{Figure2}
	\end{centering}
\end{figure}

A first experiment was performed in order to measure polarization on a single--crystal sample of CdCr$_2$O$_4$ as a function of temperature and magnetic field. The magnetic field was applied parallel to the [111] direction via a superconducting magnet. On the (111) surfaces of the sample ($A=$~16mm$^2$) we applied silver paint electrodes and measured the pyrocurrent using an electrometer (Keithley 6517). This yields the polarization by time--integration.

In figure \ref{Polarization} (a) and (b), the polarization current (pyrocurrent) and the resulting polarization at zero magnetic field and at 10~T are plotted. The data are measured by warming up the sample through the Ne\'el temperature $T_N$, at 2~K/min. In this case we did not apply a poling voltage. The magnetic field was applied both while we cooled down through T$_N$ and while we warmed up. Figure \ref{Polarization} (b) shows that at zero magnetic field no polarization is present, but at 10~T a small polarization of 0.12~$\mu$C/m$^2$ is observed below the Ne\'el temperature. Measurements were attempted with a poling voltage applied while cooling down, but there was no indication of any switchable spontaneous polarization emerging below $T_N$ at zero magnetic field. The inset in figure \ref{Polarization} (b) shows the dielectric constant, derived from the sample capacitance, measured by a capacitance bridge (Andeen-Hagerling AH2500A) at a fixed frequency of 1~kHz. A sharp decrease in dielectric constant is seen while cooling down through $T_N$.

We also measured polarization in CdCr$_2$O$_4$ at high magnetic fields, using a pulsed magnet capable of reaching 56~T. Also in this case, we maintained the same experimental conditions presented above: the electrodes were applied on the (111) surfaces of the samples, and the magnetic field applied in the [111] direction. Samples were cooled down below the Ne\'el temperature either with or without a poling voltage applied. The pyrocurrent was measured during the magnetic pulse using a high-speed digital oscilloscope. Figure \ref{Polarization} (c) and (d) shows a typical result without poling voltage. In the field increasing curve, a peak in the pyrocurrent at around 5~T signals a distinct increase of the polarization, which is related to a spin-flop transition (see below). The polarization at 10~T is comparable to that seen in Figure \ref{Polarization} (b), at 0.2~$\mu$C/m$^2$. As the field increases further, the polarization continuously increases, until at 28.3~T the polarization drops to almost zero. On the down sweep, an increase in polarization is seen at 27.5~T.

We can understand the generation of electric polarization with magnetic field in CdCr$_2$O$_4$ as being due to its spin-spiral ground state. Non-collinear spin states can produce an electric polarization either via the spin-current model \cite{Katsura2005} or an inverse Dzyaloshinskii–-Moriya-type interaction \cite{Sergienko2006}. In either case, the spin-induced polarization has the form:

\begin{equation}
P=a \sum_{<i,j>} e_{ij} \times (S_i \times S_j) 
\label{P_eqn}
\end{equation}

where  $e_{ij}$ is the unit vector connecting the neighboring spins $S_i$ and $S_j$, and $a$ is a constant which determines the coupling strength. 

Below its Ne\'el temperature, CdCr$_2$O$_4$ exhibits a spin-flop transition at around 5~T, for magnetic fields applied away from the [010] easy plane. This has been observed in magnetization and neutron diffraction data \cite{Kimura2006,Matsuda2007}, by magnetostriction \cite{rossi2019negative} and by ESR and optical spectroscopy measurements \cite{Kimura2006, Sawada2014}. We observe an increase in polarization at the spin flop (figure \ref{Polarization}(c) and (d)). This indicates a transition to a tilted conical spin structure, where the spin rotation plane $S_i \times S_j$ is canted towards the field direction, increasing the angle to $e_{ij}$ and thus per equation \ref{P_eqn}, producing an increased polarization. Above the spin flop, only a gradual increase in polarization is seen with higher fields, indicating that the spin rotation plane is close to parallel to the [111] axis by 10~T and thus that the field-induced polarization is almost saturated. At 28~T, CdCr$_2$O$_4$ enters the 'plateau' magnetic state, which is collinear and therefore generates no polarization.
 
\begin{figure*}
	\begin{centering}
    \includegraphics[scale=1]{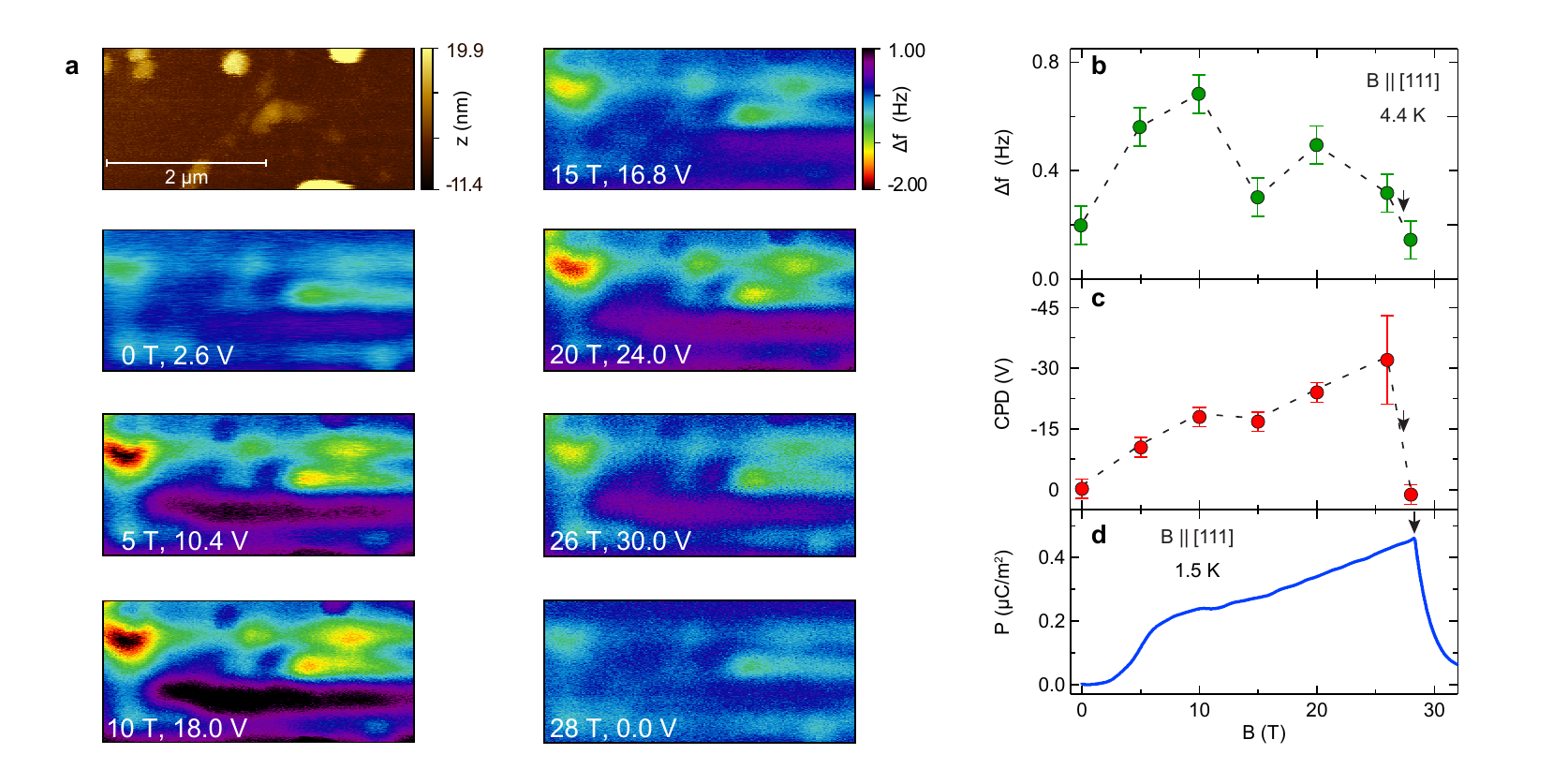}
	\caption{(a) Topographic and electrostatic force microscopy (EFM) images of CdCr$_2$O$_4$. Topographic image, top left. The EFM images correspond to the same $4\times1.8 \thinspace \mu m $ area as the topographic image, and are collected at field increments between zero and 28~T, with 100~nm lift height. The field and the contact potential difference (CPD) are indicated for each image. (b) EFM image contrast as a function of field.  The errors were calculated through repeated measurements. (c) CPD between the tip and the sample as a function of field recorded at up to 28~T at 4.4~K.  The errors were calculated through repeated measurements. (d) Magnetic field dependence of $P$ derived from pyrocurrent measurement in a pulsed magnet: P is measured parallel to the B field. A poling voltage of +500~V was applied. The transition to the plateau phase is indicated as an arrow in (b)-(d).}
	\label{Figure3}
	\end{centering}
\end{figure*}

Figure \ref{Figure2} shows the effect on the magnetic-field induced polarization of poling voltage and field orientation. It is possible to invert the sample polarization by applying an opposite poling voltage. By contrast, inverting the magnetic field orientation does not change the polarization direction. We observed that the response of the CdCr$_2$O$_4$ polarization to poling voltage produced results with a high degree of sample dependence. In some samples, for example, it was not possible to invert the polarization, but the polarization direction was apparently ``pinned''. We observed also some history dependence, in terms of the magnitude of the polarization obtained. We attribute these effects to the presence of structural domains in the low-temperature tetragonal antiferromagnetic phase. Different orientations of domains affect the total polarization, since the spin spiral vector is locked to the tetragonal lattice. Additionally, domain walls can act as charge trapping sites. Since domains are erased every time T increases above 8~K, there is a degree of stochasticity in the pattern of domains created each time. This may account for the difference between field-increasing and -decreasing plots seen in figure \ref{Polarization}, since the domains are also erased on the field-induced transition. This sample- and history- dependence made more detailed studies of the total polarization difficult, for example its field-orientation dependence. 

A way to avoid the problems associated with domain structures is to perform local polarization measurements within an area much smaller than the size of the domain. To this end, we performed field dependent electrostatic force microscopy (EFM), on the as--grown (111) surface of CdCr$_2$O$_4$ single crystals at a range of static fields up to 28~T. We also recorded the contact potential difference (CPD) between the tip and the sample at each field. The CPD is the bias, applied to the sample, that is needed to nullify the electrostatic forces, measured as phase or frequency shift, between the tip and the sample. This technique is normally used to compensate for a work--function difference which is typically $\sim1~V$, but in this experiment we recorded the CPD voltage in order to measure a surface charge induced by bulk polarization. The CPD was measured at a single point in the scan area, before recording the image, and it was kept constant during the scan. The CPD was measured with the SPM tip at a fixed height of 100~nm above the sample surface. By applying a bias voltage to the sample via a silver paste connection, we could measure the voltage which is induced at the surface of the sample, due to the bulk polarization. For this purpose, an SPM able to work in high magnetic field was employed \cite{Rossi2018}. We operated the HF--SPM at 4.4~K, in He--exchange gas at a pressure of around $10^{-2}$~mbar. All the data presented were recorded using a Co--coated cantilever with 300~kHz resonance frequency and a Q--factor of 3800 at zero field. The images were collected in amplitude modulation mode (AM--AFM or tapping mode) at a fixed frequency of 320~kHz. EFM images and CPD voltage were collected in lift mode with 100~nm of lift height. The applied magnetic field was parallel to the [111] direction of a CdCr$_2$O$_4$ crystal.

Figure \ref{Figure3} (a) shows topographic and EFM images of CdCr$_2$O$_4$. In order to obtain the EFM images we recorded the phase shift image at field increments between zero and 28~T. Since the Q--factor changed with field we translated the phase measurements into frequency shift. We then extracted the frequency shift values calculating the RMS image contrast and plotted it as a function of field, see figure \ref{Figure3} (b). No qualitative change in the EFM images is seen with field. However, the image contrast does change, being low at 28~T and zero field, and higher at intermediate fields. We see no indication of tetragonal domain walls in the images of figure \ref{Figure3} (a), which in the observed [111] orientation would be expected to present linear features along 120$^{\circ}$ domain boundaries. This indicates that the size of the domains is larger than the scan area.  

Figure \ref{Figure3} (c) presents the CPD voltage as a function of field. Normally we would not expect the CPD to vary with magnetic field $B$, but here we recorded a clear change in the CPD with $B$, which increases with field up to 26~T and then goes back to zero. The recorded CPD values are also strongly enhanced, reaching up to 30~V, compared to typical CPD values due to work--function differences of $\sim1~V$.  
This curve can be compared to polarization data derived from pyrocurrent from a similar sample, in a pulsed field at 1.5~K, figure \ref{Figure3} (d). Both data sets show a continuously increasing signal in polarization or bias up to the the transition to the plateau phase. Once the plateau phase is reached, the polarization and bias both drop back to zero. The magnetostructural transition occurs at different fields between figure \ref{Figure3} (c) and (d) due to the temperature dependence of the transition: \cite{rossi2019negative}. To identify precisely the magnetostructural transition field in the SPM experiment we used the HF--SPM as a dilatometer \cite{Rossi2018}.  From this comparison we can conclude that CPD voltage measurements with SPM are equivalent to polarization.

The measurement of CPD in figure \ref{Figure3}(c) provides a single point measurement of the bulk polarization. We can also investigate the surface charge distribution via the EFM images in \ref{Figure3}(a). The EFM images do not resemble the topography, but indicate an inhomogeneous distribution of surface charges, likely due to the presence of defects at or near the surface. The pattern of defects does not change with magnetic field, but as the polarization increases with field, the presence of defects leads to an increasing charge inhomogeneity and hence increased image contrast. The effect of the defects is apparently less pronounced at high fields, leading to a diminished EFM image contrast above 10~T (figure \ref{Figure3}(b)), whereas the CPD and polarization peak just before the transition at 28~T. At the transition to the plateau state, both the CPD and image contrast revert to their values at zero magnetic field.
 
To conclude, we have observed electric polarization induced by magnetic field in the frustrated spinel CdCr$_2$O$_4$, for the first time. Polarization arises due to the non-collinear spin spiral state. We see no evidence for spontaneous polarization at zero magnetic field. Also the sign of polarization does not change with magnetic field direction, but only with poling voltage sign. Thus in terms of the electric polarization generated via equation \ref{P_eqn}, the spin-spiral ground state of CdCr$_2$O$_4$ behaves like a simple proper screw, as per ZnCr$_2$Se$_4$ \cite{Murakawa2008}, and not like a spin cycloid as in CoCr$_2$O$_4$ \cite{Tokura2010, Yamasaki2006}. This is despite CdCr$_2$O$_4$ having in fact a somewhat more complex incommensurate spin spiral \cite{Chung2005, Matsuda2007, Chung2013}.

In polarization, we observe clearly the effect of the spin flop transition at $\approx$ 5~T. Based on ESR and optical spectroscopy measurements \cite{Kimura2006, Sawada2014} the spin-flop has been interpreted as a transition from a helical structure to a commensurate canted spin structure. This would imply the disappearance of polarization above 5~T, which is not observed in our polarization data. Therefore, we can conclude that a spin spiral state survives as a canted conical spiral up to the transition to the collinear half-magnetization plateau at 28~T. This conclusion is supported by neutron diffraction experiments \cite{Matsuda2007}.

We have demonstrated that high magnetic field electrostatic force microscopy provides a viable alternative way of measuring electric polarization. By enabling polarization measurements to be made on a nanometer length scale, it was possible in this study to measure polarization within domains. The technique could readily be improved by the addition of Kelvin probe feedback: this would allow to record CPD at each pixel whilst imaging, or to make continuous measurements of CPD while sweeping the field. This new method for measuring polarization will be a valuable tool for studying multiferroic samples with nano-and meso- scale inhomogeneities such as multiferroic domain walls \cite{Meier2015} and engineered nanostructures \cite{Martin2008, Chen2016, Zhong2019}. 

\begin{acknowledgments}
The authors would like to thank Jan Gerritsen for valuable assistance with the SPM measurements. D. Br\"uning and T. Lorenz acknowledge support by the DFG (German Research Foundation) via Project No. 277146847-CRC 1238 (Subproject B01).
\end{acknowledgments}

\bibliography{Bibl}

\end{document}